# Anvil cirrus suppresses tropical altocumulus

Wouter Mol[1], Blaz Gasparini[1], Aiko Voigt[1,2]

[1]*Department of Meteorology and Geophysics, University of Vienna, Austria*
[2]*Institute of Geosciences, University of Bonn, Germany*

## Key points

- Altocumulus, a stratiform mid-level cloud, is common around tropical deep convection and is frequently overlapped by cirrus
- Cloud resolving simulations show anvil cirrus weakens or dissipates altocumulus
- Altocumulus does not impact mid-level stability in a way that explains or reinforces its own formation

## Abstract

Tropical cloudiness in convective regions has three modes: boundary layer, mid-tropospheric, and cirrus clouds. In this study, we explore tropical cloudiness with a focus on the role of mid-tropospheric altocumulus and its formation. First, we use satellite data to show that multi-layered cloudiness is more common than single-layer, with altocumulus and cirrus making up 20% of all two-layered cloud systems. Second, we design high-resolution numerical experiments that simulate these radiatively coupled altocumulus and cirrus. The experiments show altocumulus lifetime is halved by thick cirrus. Most altocumulus is therefore unlikely to form directly from detrained deep convection, as it would be suppressed by anvil cirrus, and instead favors congestus detrainment or large-scale ascent as mechanisms. By the same argument, changes in anvil cloud cover will affect mid-level cloudiness and thus tropical climate. Given the high resolution required, this connected cloud layer perspective is not captured by climate models.

## Plain language summary

Cloudiness in the middle and upper part of the tropical atmosphere is almost solely related to moisture originating from thunderstorms. The high clouds that are formed as part of these thunderstorms, known as anvil cirrus, get much attention due to their significant but uncertain role in climate. Yet, anywhere in these regions with active storms we also find a maximum in clouds around 5 km altitude, from so-called altocumulus clouds. The origin, maintenance, and role of these clouds in climate are not well understood and therefore are the focus of this study. We show using satellite observations that altocumulus often overlaps with cirrus clouds. With computer simulations that resolve the small-scale cloud dynamics and effects of solar and thermal radiation, we show that the thicker the cirrus clouds are, the more the altocumulus thins. This happens through increasing downwelling thermal radiation with increasing cirrus thickness. Our results also suggest altocumulus likely originate at a slight distance from the thunderstorms, rather than directly under the thickest anvils. Understanding the origins and connections between separate cloud layers further helps understand current tropical cloudiness and changes therein due to climate change.

## 1 Introduction

Low and high clouds have a prominent place in the atmospheric and climate science community due to their clear and significant radiative effects. For clouds in the mid troposphere, this picture is less clear, yet they are abundant (Zhang et al. 2010; Bourgeois et al. 2016; Stevens et al. 2026), interact with radiation and atmospheric circulation (Sokol and Hartmann 2022; Spaulding-Astudillo and Mitchell 2025), and frequently dampen the radiative effects of low or high clouds by overlapping with either (Pan et al. 2025).

Mid-level clouds are most prominently featured in the context of tropical cloudiness, where regions of convective activity have long been known to feature a trimodal cloud distribution with height (Johnson et al. 1999). The likely explanation is a weak temperature inversion around the 0 degree isotherm, which enhances detrainment of convective updrafts at that height (Spaulding-Astudillo and Mitchell 2025), but this is not captured in models that describe tropical cloud cover as a function of convective detrainment and cloud lifetime (e.g. Seeley et al. 2019). Why there is a temperature inversion in the first place is still actively discussed (Sokol and Hartmann 2022; Spaulding-Astudillo and Mitchell 2025).

We first have to clarify the different main types of mid-level clouds in the tropics. Cumulus congestus, rooted in the boundary layer and with cloud tops up to approximately 9 km, are the optically thick cloud type discussed by Johnson et al. (1999). However, congestus updrafts are small, so it is the stratiform and shallow (<1 km) altocumulus and elevated convection (approx. 1-5 km deep) that dominate the cloud cover climatology. Precise numbers depend on cloud definitions, measurement type, instrument sensitivities, and spatiotemporal coverage. The global yearly average mid-level cloud cover is approximately 10 to 20% (Stubenrauch et al. 2013; Luo et al. 2023), while 6-8% of cloud cover can be attributed specifically to altocumulus using remote sensing data (Sassen and Wang 2008; Zhang et al. 2010), or 17% based on surface observations (Hahn et al. 2001).

Seasonal, regional, and daily variations are substantial. Based on remote sensing data, in the tropics in and around convectively active regions the altocumulus cover increases from 6-8% to over 15% daily averaged, with regionally as much as 4 to 10 percentage points more during night than daytime (Zhang et al. 2010; Riihimaki et al. 2012). The proximity of altocumulus to deep convection means overlap with (anvil) cirrus is the most common type of overlap in the tropics (Li et al. 2015). A similar result is found for "high" and "middle" cloud overlap by Liu et al. (2025).

The reason the knowledge of mid-level clouds has mostly come from observational studies is likely because most models cannot properly represent altocumulus. These stratiform mixed-phase clouds are typically only several hundred meters deep, have (supercooled) liquid tops, and feature small-scale internal convective cells. This makes them similar to stratocumulus, except they are disconnected from surface heat fluxes. Most models have much less resolution in the mid troposphere compared to the boundary layer, such that all that makes altocumulus interesting is unresolved and fully parameterized. Large-eddy simulation studies, where these clouds are explicitly resolved, show altocumulus lifetime is sensitive to the longwave cloud top cooling, which is the one mechanism prolonging lifetime by compensating for moisture removal by precipitation, subsidence, and entrainment of dry air (Larson et al. 2006).

In this paper we aim to clarify the role of mid-level clouds for tropical cloudiness, with a particular focus on the role of cloud overlap. Since the altocumulus (hereafter "Ac") are expected to frequently overlap with (anvil) cirrus (hereafter "Cs") and both layers alter radiative fluxes in the atmosphere, we expect Ac to dissipate faster in the presence of Cs (due to reduced longwave top cooling) and Cs to persist for longer (due to reduced heating from upwelling longwave radiation). Our overarching hypothesis is that accurately representing this radiative interaction between cloud layers is necessary for reproducing observed tropical cloudiness. Section 2 is an observation-focused analysis which sets the foundation for the numerical experiments in Section 3. Building on those results, we touch upon the question of whether Ac impacts its own formation in Section 4.

# 2 Mid-level cloud climatology

## 2.1 Remote-sensing data

We use CloudSat-CALIPSO data for estimating the vertical cloud climatology in specific regions in the tropics. Specifically, we use the `2B-FLXHR-LIDAR.P2_R05`, `2B-CLDCLASS-LIDAR.P1_R05`, and `ECMWF-AUX.P1_R05` products available from cloudsat.cira.colostate.edu. These products are combined retrievals of the low-Earth-orbiting satellites CloudSat (carrying a cloud profiling radar, Tanelli et al. (2008)) and CALIPSO (carrying a cloud-aerosol lidar, Winker et al. (2010)). The `2B-FLXHR-LIDAR.P2_R05` retrieval product provides a vertically resolved cloud optical depth profile with 240 m grid spacing and an approximate 1.8 km along-track and 1.3 km cross-track resolution. The `2B-CLDCLASS-LIDAR.P1_R05` is an algorithmically derived product where individual cloud layers are detected and categorized. Co-located profiles of the ECMWF's atmospheric analysis (`ECMWF-AUX.P1_R05`) provide temperature and humidity data.

The low orbit and narrow horizontal footprint of these satellites are such that the revisit time of a single location is approximately 16 days (Tanelli et al. 2008), meaning we need to average sparse spatiotemporal snapshots over a larger region to derive statistics. Based on previous work (Zhang et al. 2010; Bourgeois et al. 2016; Stevens et al. 2026), we choose a region encompassing the Sahel in West Africa as our primary interest (-20 to 20E, 0 to 30N). Other convective hotspots in the tropics, Amazonia (-80 to −40E, −20 to 10N) and the West Pacific Warm Pool (90 to 160E, −15 to 15N), are added to show the Sahel is not unique in showing mid-level cloudiness. We primarily take data of the year 2010 which we found to be the most recent and most complete year with both daytime and nighttime data. Results from the second-most complete year, 2007, are qualitatively similar.

We use the Ac and Cs classes from the cloud classification product. These are broadly representative of the cloud type of our interest in this study. While for cloud overlap statistics we use all data, we only consider profiles with 1 or 2 cloud layers, containing Ac and/or Cs. For Ac, the classification makes no distinction between true Ac (stratiform liquid-topped mixed-phase cloud) and elevated convection (vertical depth of several km), but they are both wider-than-tall cloud fields with internal convection due to top cooling and bottom warming. The spatial scales of this deeper kind of Ac go beyond what we represent in our cloud-resolving model experiments (Section 3). We use broadband longwave optical depths as these are available for both daytime and nighttime data.

## 2.2 Trimodal cloud fraction and cloud overlap climatology

Figure 1 shows the vertical cloud climatology and cloud overlap statistics for three convectively active regions in the tropics. Over the West African Sahel region during the summer monsoon, we find the three expected modes of cloudiness. The mid-level mode is most pronounced for the lower optical thickness thresholds ($\tau_{\mathrm{lw}} \leq 0.1$ layer$^{-1}$, one layer is 240 m), suggesting it is predominantly the shallow altocumulus contributing to the cloud cover rather than congestus. Yearly-averaged data for the Amazonia region (Figure 1b) and West Pacific Warm Pool (Figure 1c) show overall higher cloud cover, but are otherwise qualitatively similar.

The local maxima in cloud cover at 260 K for the highest threshold (1 layer$^{-1}$) likely contains mostly the deeper, elevated convection mode of mid-level cloudiness, whereas the jump in thick cloud cover below the freezing level is due to the transition from frozen to liquid precipitation (since precipitation is included in the satellite retrievals). Based on the difference between the 1 and 0.1 layer$^{-1}$ thresholds, most thin mid-level clouds (Ac) thus sit closer to the 0 degree isotherm.

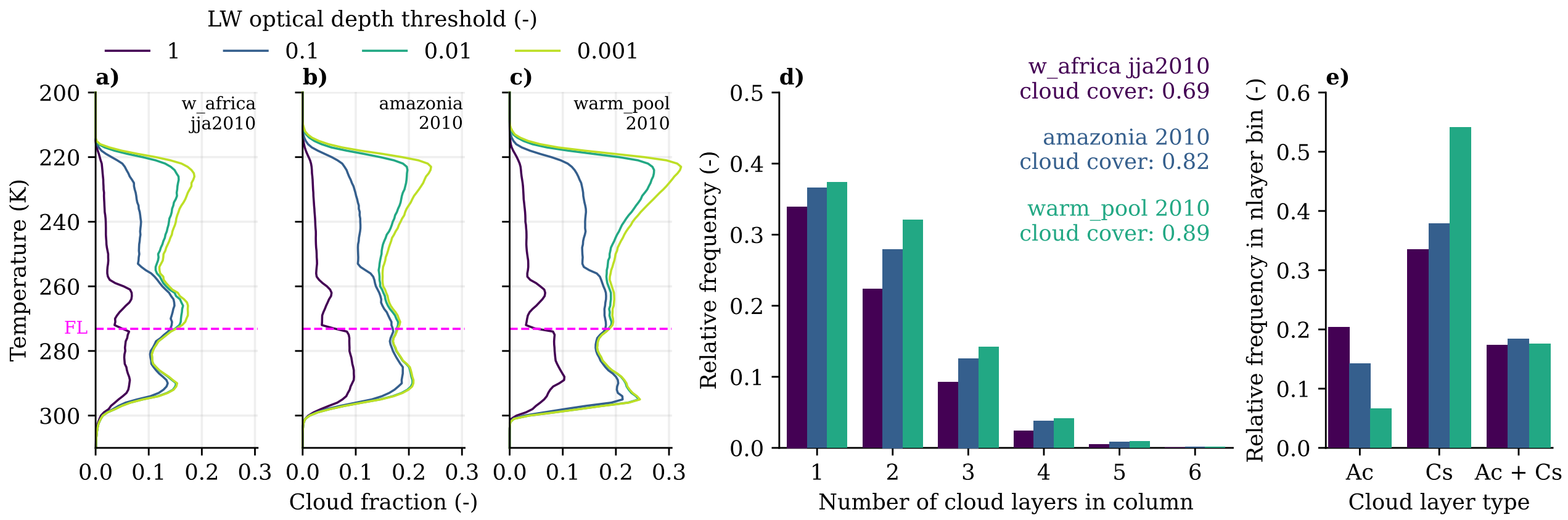

Figure 1: **Vertical cloud climatology derived from CloudSat-CALIPSO retrievals**. Subplots **a-c** show for four thresholds of longwave optical depth per layer (= 240 m depth) a seasonal or yearly average of vertical cloud fraction for West Africa (**a**), Amazonia (**b**), and West Pacific Warm Pool (**c**) regions, regridded to temperature coordinates using co-located ECMWF data. For the same regions and temporal ranges, subplot **d** shows the frequency of cloud overlap occurrence and **e** quantifies the contribution of Ac, Cs, and Ac+Cs within the 1, 1, and 2 layer bins, respectively.

Cloud overlap is more common than single layer cloud cover in these three regions (Figure 1d). The Cs and Ac types (Figure 1e) reveal that, together, these cloud types make up over half of single layered cloud conditions. For two-layered cloud conditions, the combination of Ac and Cs accounts for close to 20% of cloud cover. Interestingly, single-layer Ac becomes less common with increasing Cs cover, while the relative occurrence of Ac + Cs overlap stays constant across the three regions.

Satellite retrieval statistics support weaker altocumulus in the presence of optically thick cirrus ($\tau_{\mathrm{lw}} \geq 0.5$, Figure S1). The median optical depth of the Ac layers, here filtered to be no deeper than 1 km (4 grid cells), is lower by $\tau_{\mathrm{lw}} \approx 0.3$ in each of the three regions. However, optically thick cirrus can obscure layers below it by attenuating the lidar signal, making non-precipitating Ac undetectable to both lidar and radar. We also find the median reduction can vanish in one or more regions for different combinations of thresholds for thin/thick cirrus ($\tau$ = 0.3 to 1) and Ac vertical depth (3 to 5 grid cells). Lastly, if thick Cs dissipates Ac, the statistics are subject to survivorship bias, as we cannot observe Ac that would have been there had they not been dissipated due to the Cs's presence. For more robust answers to our question of cloud overlap impacts, we now turn to numerical experiments.

# 3 Numerical experiment design and results

## 3.1 Cloud-resolving large-eddy simulation

We run idealized cloud-resolving atmospheric simulations using MicroHH 2.0, a large-eddy simulation model (van Heerwaarden et al. 2017). The general idea of the experiments is that we force the formation of one or two cloud layers in the first hour and then let the simulation run freely where only local radiative and microphysical processes act to dissipate (or invigorate) the cloud layer(s).

We target the cloud layers in our numerical experiment to match composite statistics of geometry and optical depth from the satellite data over West Africa, though with some adjustments. The altocumulus is modified to be shallower than this climatology, as the altocumulus cloud class also includes elevated convection that cannot be sustained in small-domain simulations. The two cloud layers are therefore located initially at 5200 - 5400 m and 12400 - 13900 m for Ac and Cs, respectively.

The simulations are initialized using a mean profile of the ERA5 temperature and humidity over the Sahel for July 2020 (Figure 2). The target cloud layer heights are set to 100% relative humidity (with respect to water), mixed to be neutrally buoyant, and with a 1 K (Cs) or 3 K (Ac) capping inversion to limit immediate detrainment. Then, there is a one hour spin-up phase in which we force the formation of a cloud layer by prescribing a tendency to the prognostic variables $q_t$ (specific humidity minus hydrometeors) and $\theta_l$ (liquid water potential temperature), letting the model cloud dynamics and microphysical processes adjust.

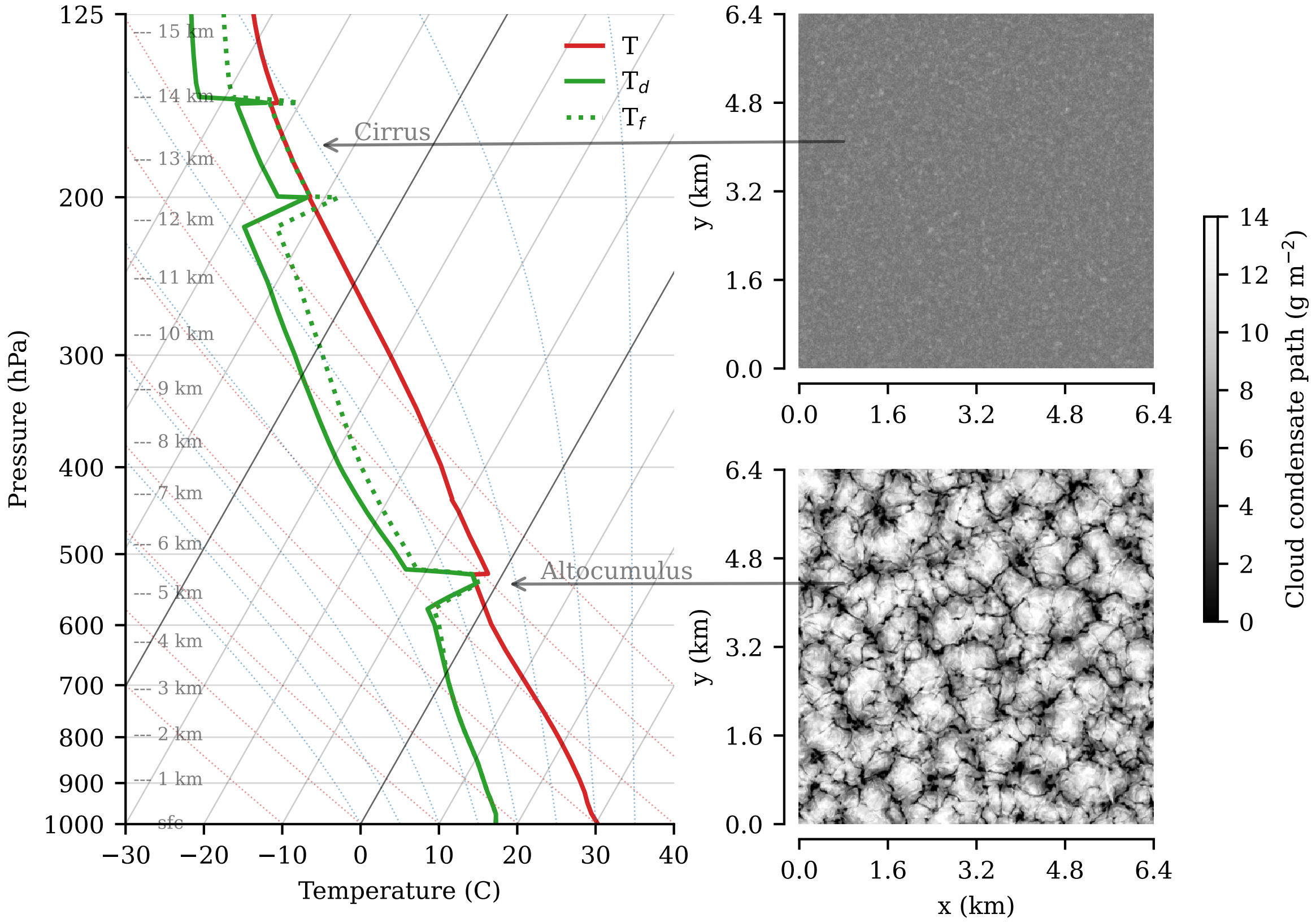

Figure 2: **Numerical simulation setup**. The modified Sahel monthly mean profile of temperature (T), dew point ($T_d$), and frost point ($T_f$), used as initial conditions, are shown in a skew-T log-P diagram. Vertically integrated liquid and frozen cloud water for Cirrus and Altocumulus are shown on the right for snapshots 1 hour after spin-up.

The tendencies are chosen such that the initialization results in 10 g m$^{-2}$ ($\tau \approx 3$) and 25 g m$^{-2}$ ($\tau \approx 1$) of vertically integrated cloud condensate for Ac and Cs, respectively. This matches the highest end of the satellite retrieval data that we can reasonably simulate. The reason is that if we see an effect from the cloud overlap, it should be most clear for the conditions in which the radiative fluxes are influenced the most. For the Cs layer, we add a quasi-fixed Cs layer variant that keeps the tendencies fixed throughout the simulation, meant to represent fresh anvil formation. Optical depths $\tau$ for the simulations refer to broadband shortwave radiation.

The spatial resolution required to represent the stratiform mid-level clouds dominates our numerical design choices. The cloud layer requires at least two grid cells to resolve a radiative heating contrast, setting the maximum reasonable vertical grid spacing to 50 m, which accounts for Ac becoming shallower as it dissipates. Internal convection cells have scales of $\sim$ 100 m, meaning the horizontal grid spacing has a similar maximum. Sensitivity experiments to test domain size, horizontal, and vertical resolution suggest that at a 6.4 x 6.4 x 15.4 km$^3$ domain with $\Delta x, y, z$ = (15, 15, 16.7) m grid spacing the domain mean vertically integrated condensate and top-of-atmosphere radiative balance have mostly converged; all experiments are therefore run on this grid.

Microphysics is kept simple and computationally efficient, and thus requires no tuning: we use a single-moment scheme where liquid and frozen condensate are diagnosed based on excess water vapor and prognostic variables are snow, graupel, and rain. Partitioning between liquid and frozen condensate is linear between 0 °C (all liquid) and −40 °C (all ice). This correctly makes the Ac liquid-dominated mixed-phase and thus has the necessary cloud top cooling. The Cs layer is 100% ice, but cloud ice sublimates instantly in response to reduced supersaturation, which we suspect shortens cirrus lifetime. We deal with this misrepresentation by adding an experiment where the cirrus layer is forced to persist through the simulation. The number concentrations are kept fixed at 2.5e8 m$^{-3}$ (liquid) and 1e5 m$^{-3}$ (ice).

For online radiative transfer calculations, MicroHH uses RTE+RRTMG-CPP, which calculates radiative fluxes using the two-stream approximation, here set to update every 30 seconds. For offline calculations of optical depths and radiative fluxes from 3D output we use ecRAD (Hogan and Bozzo 2018). For the rest, horizontal wind is set to 0, the lateral boundaries are periodic, there is no daily cycle or interactive land surface, large-scale vertical motion is set to zero, and aerosols are not included.

## 3.2 Anvil cirrus shortens altocumulus lifetime

Figure 3 shows the evolution of total cloud layer condensate and radiative heating for the four experiments we run: Ac-only, Ac with Cs, Cs-only, and Ac with a quasi-fixed Cs layer. The Ac layer in our model starts precipitating early on in the spin-up phase (t = −0.5 h), which comes into balance with the constant $q_t$ and $\theta_l$ tendencies at t = 0 h. The remaining 12 hours show a slowly dissipating Ac layer in which sedimentation and dry air entrainment are largely offset by layer-mean cooling. Convection within cloud layers is visible as temporal variability in the domain mean condensate, due to the relatively small domains used. Stronger radiative cooling also results in stronger within-cloud heterogeneity (see Figure 2 for a typical snapshot of 2D cloud structures in Cs and Ac).

In contrast to Ac, the Cs layer sedimentation rate of ice is too high to sustain the Cs cloud at $\tau \approx 1$. The simulation with "fixed" Cs also varies over time due to internal cloud dynamics, though more weakly than Ac.

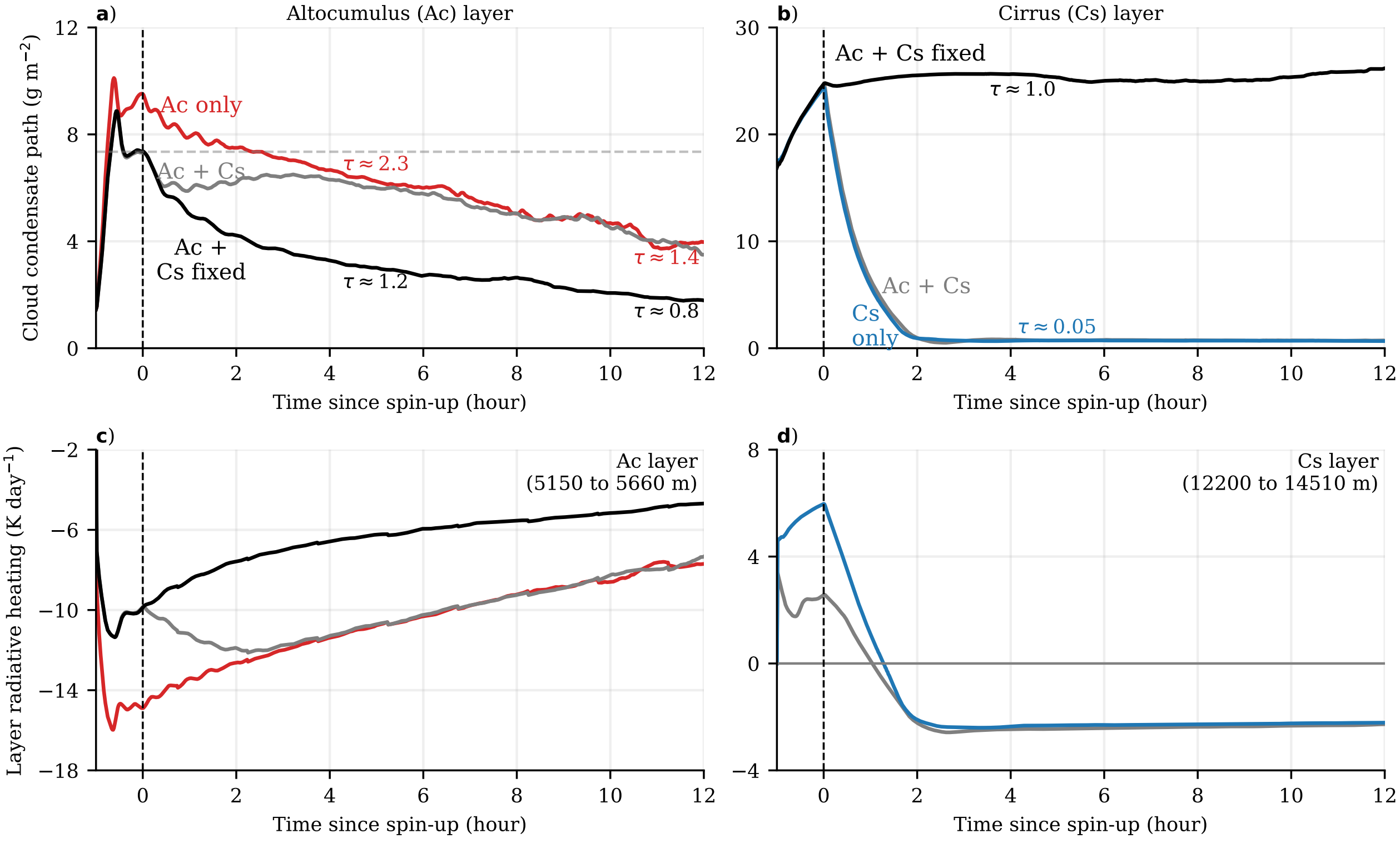

Figure 3: **Time series of simulation output** for Altocumulus (left column, **a,c**) and Cirrus (right column, **b,d**) cloud layers of the night-only Ac, Cs, Ac + Cs, and Ac + fixed Cs experiments. Layer radiative heating is a vertical mean that closely follows the cloud layer and is converted to units of K day$^{-1}$.

This main experiment primarily shows how the optical depth of the Ac layer, and thus its effects on the environment and Cs, depend on how effectively it can radiatively cool. The Ac-only experiment simulates a slowly dissipating Ac with $\tau \approx 2.3$ at t = 5 h. The same Ac but with a cirrus layer above starts at a lower optical depth due to already significantly reduced cooling from the overlying Cs during the spin-up, reaching $\tau \approx$ 1.2 at t = 5 h in case the Cs is kept quasi-fixed. The difference in downwelling longwave radiation between these two conditions is 15 to 20 W m$^{-2}$. In case the cirrus dissipates, however, the radiative cooling recovers within 2 hours and the Ac becomes as optically thick as the Ac-only case in 4 to 5 hours. For a given layer's cloud condensate path ($Q = \int (q_i + q_l) \rho \, \mathrm{d}z$), the dissipation rate of the Ac ($\partial Q / \partial t$) with the quasi-fixed Cs is consistently higher, meaning the decreased lifetime is not due to a difference in spin-up.

The Cs appears largely unaffected by the presence of Ac. Until $t = 1$ h, the net radiative heating rate throughout the layer is less than half of that with Ac present (2 vs. 6 K day$^{-1}$). This reduction in warming is not enough to meaningfully compensate the loss of moisture by sedimentation before the Cs all but dissipates. After 2 hours, the Cs layer is a stable non-precipitating cloud of $\tau \approx 0.05$, such that the layer has no more radiative or microphysical impact on the atmosphere or the Ac layer at these time scales. Nothing significantly changed in sensitivity tests in which the Ac was kept fixed or made more optically thick.

Increased robustness of Ac with increased cloud-top longwave radiative cooling becomes more apparent when we add solar radiation (Figure S2) or subsidence (Figure S3). Adding constant peak daytime solar radiation completely dissipates the Ac with a quasi-fixed Cs in 3 hours, while without Cs the Ac layer is still present with 1 g m$^{-2}$ at $t = 6$ h. Reduced solar irradiance in the presence of Cs therefore does not offset the reduced longwave cooling for the Ac. Adding subsidence, equivalent to a rate at which the layer rises during the simulation due to entrainment ($w \approx 0.65$ cm s$^{-1}$), more than halves the lifetime of Ac in case Cs is present (1.5 h instead of 4 h). For the daytime experiment, the Ac can still recover in time when the Cs dissipates, but in the case of subsidence (but no solar irradiance) the weakened Ac does not recover and dissipates in 2.5 h, still faster than in the Cs-free case.

Our idealized tropical altocumulus is more sensitive to these environmental factors, subsidence in particular, compared to what Larson et al. (2006) show, likely because their case study features a deeper (500 m), more continuous "altostratocumulus".

# 4 Discussion

## 4.1 On the origin of tropical altocumulus

We now come back to the question of Ac formation, mid-level stability, and whether mid-level cloudiness strengthens this stability and thus reinforces its own formation. Our small, periodic domain experiments limit making strong conclusions about this question, so we intend to offer the following two thoughts as hypotheses.

First, tropical Ac is more likely to originate from congestus detrainment or the later lifting of detrained moisture, rather than directly from detrained deep convective clouds. Limited longwave cooling due to freshly generated anvil Cs and subsidence surrounding strong updrafts will not support long-lived Ac, as per our experiment results. If later ascent in the moist layer comes from convectively generated gravity waves, we expect that Ac, once formed, can resist complete evaporation from the subsiding part of the wave owing to longwave cooling. This requires the overlying cirrus to have thinned, meaning either the deep convection has subsided or the moist layer has been advected away to a calmer part of the atmosphere. Should the convective mass detrainment come from congestus clouds that terminate in the mid troposphere as discussed by Sokol and Hartmann (2022), Ac can also persist.

Second, Ac is unlikely to influence renewed Ac formation. Figure 4 shows the net cooling rate and stability change after 12 hours of nighttime conditions of Ac compared to cloud-free conditions. The temperature anomaly is allowed to build up in these simulations due to the horizontally uniform domain, whereas in reality the diabatic heating would result in a circulation. We can estimate the circulation from the diabatic heating ($\Delta\theta/\Delta t$) and vertical temperature gradient ($\partial\theta/\partial z$), which suggests weak ascent in the lower 5 km ($\approx 0.2$ cm s$^{-1}$) and descent near cloud top. Whether the resulting horizontal circulation is beneficial to more convection depends on where the Ac layer is with respect to convective activity and whether the near-surface flow draws in a meaningful amount of relatively dry or moist air. In regards to the mid-level stability (Figure 4b), we diagnose strong local changes but a weak mean decrease. This may focus detrainment for new convection, but only in case such convective updrafts do not immediately mix these layers, which we deem unlikely.

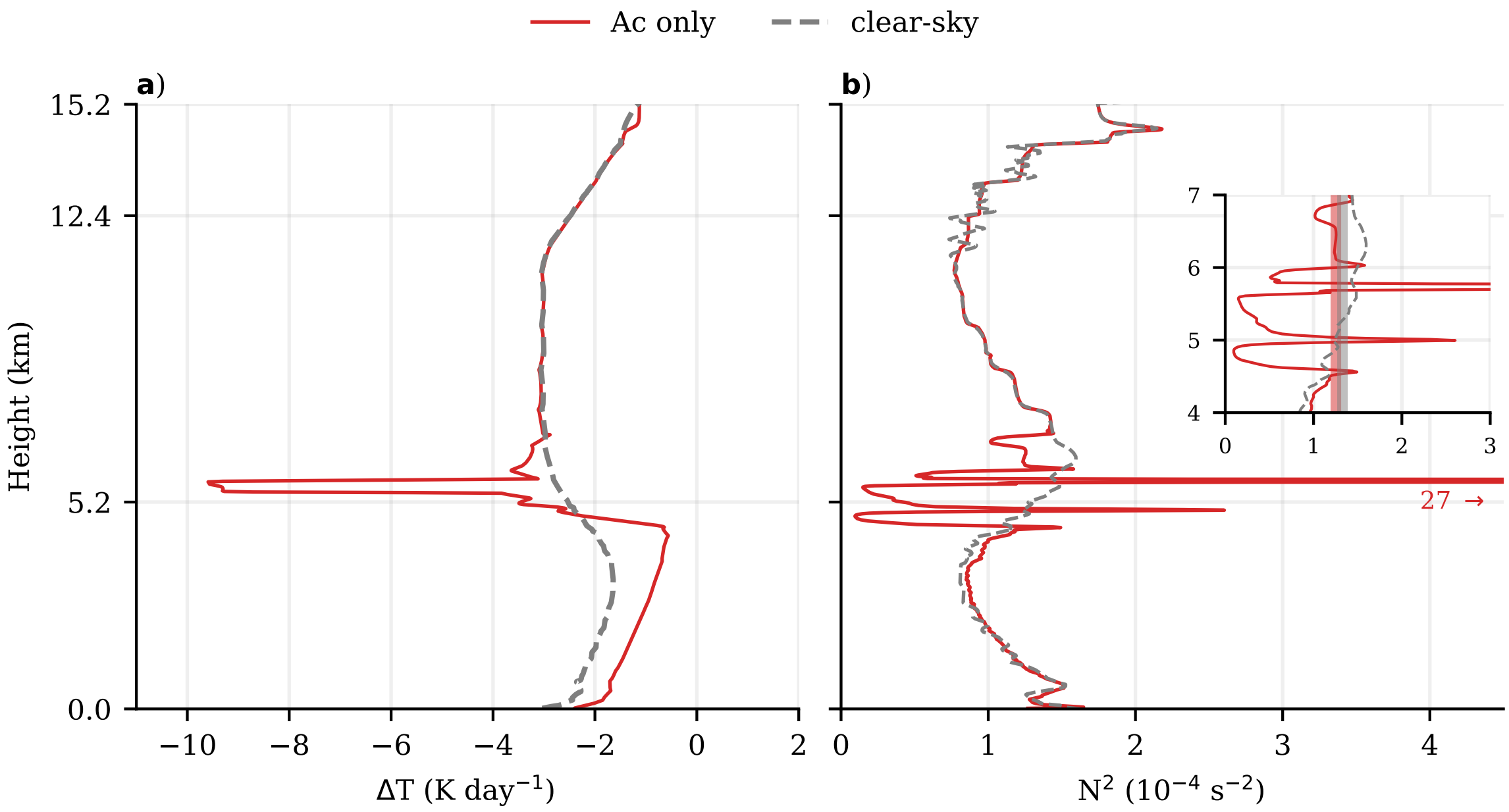

Figure 4: **Vertical profiles of temperature and stability change** in the Ac and clear-sky simulations. **a** shows the temperature change averaged over hours 11-12 compared to hours 0-1, converted to K day$^{-1}$. **b** shows the mean stability over hours 11-12. The vertical red and gray shading in the zoom inset are the mean stabilities over the 4-7 km range.

## 4.2 The need for resolving altocumulus clouds in climate models

Finally, we discuss what Ac + Cs overlap means for the top of atmosphere cloud-radiative effect (TOA CRE), relevant for climate. One pathway that overlap has an impact is by Cs weakening the Ac, meaning in cases of Ac + Cs overlap the combined radiative effect is less than if the Ac were not weakened by Cs. From offline radiative transfer calculations, we estimate that effect to be only 2 to 5 W m$^{-2}$ at TOA. This does not amount to anything significant on the global scale after factoring in the low total cloud fraction of thick anvil.

Two other pathways are potentially impactful, however, and these motivate the need for accurate representation of Ac in climate models. First, the longwave TOA CRE of Ac in our simulations is $\sim$30 W m$^{-2}$, meaning this is an extra amount of longwave radiation lost to space when Ac is completely dissipated by Cs and unable to re-form later. Underresolving Ac and, by extension, misrepresenting the response to Cs overlap can thus introduce biases in simulated Ac cloud cover climatology.

Second, consider a scenario in which anvil coverage changes in a warmer climate (Bony et al. 2015; McGraw et al. 2026). If high level cloudiness in the tropics reduces irrespective of mid-level detrainment, less Ac will be dissipated by anvil clouds and the overall robustness and coverage of Ac will increase. During nighttime, Ac would thus dampen the effect of increased TOA outgoing longwave from a reduction in Cs, whereas during daytime the shortwave effect dominates and Ac will amplify the radiative effect of a reduction in Cs. Since Ac persists for longest during nighttime conditions, we expect this dampening effect to dominate, i.e., Ac could offset part of a potential negative warming feedback of anvil cirrus.

# 5 Conclusion

We have demonstrated how overlapping cloud layers can meaningfully impact each other through altering the up and downwelling radiative fluxes. Our focus was on tropical mid and high level clouds, common near tropical convection. We have shown how altocumulus in particular is sensitive to how much it can radiatively cool, a cooling strength which is controlled by overlying cirrus. Altocumulus lifetime in conditions of weak subsidence or daytime solar heating is halved by the presence of anvil cirrus ($\tau \sim 1$), but (partial) recovery is possible if the cirrus becomes optically thin ($\tau < 0.1$).

The cirrus lifetime was not impacted by the altocumulus, though we are less certain of this result, in part due to the simple ice microphysics but more so due to the horizontally uniform and periodic domain, which inhibits the formation of circulations larger than the cloud-scale turbulence. Allowing such circulations is relevant for the spreading rate and lifetime of anvil cirrus (Sullivan and Gasparini 2024; Lutsko et al. 2026) and Ac would possibly also evolve differently.

Of course, there are many other processes at play that can change our estimates of cloud lifetime derived from treating cloud overlap in an idealized way. Our ideas on altocumulus formation and their role in the tropical atmosphere should be tested in cloud-resolving simulations for domain sizes that can embed multi-day deep convection and where Ac can form directly from detrained convection or indirectly from lifting elevated moist layers by large scale, mesoscale, or gravity wave processes. We estimate this would require a domain size of at least 100 x 100 km and little compromise on resolution, meaning approximately 250x the amount of GPU resources (∼6 TiB video memory in our case). A clever way of grid-nesting, with a fine mid-tropospheric grid within a larger mesoscale grid, may offer some near-future perspective on making this possible with fewer computational resources.

We thus put three ideas on Ac formation forth as hypotheses rather than conclusions. First, Ac more likely originates from congestus detrainment or from lifted elevated moist layers than directly from deep convection, due to anvil cirrus' radiative effects. Second, and by the same argument, a reduction in high-cloud cover increases mid-level cloudiness in the tropics. Third, Ac contributes little to its own formation, since we found no evidence for increased mid-level stability. Common to all three is our view that tropical cloudiness can only be understood by treating multi-layered clouds as a connected system, in which shallow stratiform clouds are a particularly sensitive component.

# Open Research

Scripts to process CloudSat-CALIPSO data, perform the model runs, all data necessary for producing the figures, and the scripts that produce the figures are available on Zenodo: https://doi.org/10.5281/zenodo.24492508. CloudSat-CALIPSO data were downloaded from https://www.cloudsat.cira.colostate.edu/. ERA5 data were downloaded from https://cds.climate.copernicus.eu/.

# Acknowledgements

WM conceptualized the study, performed the analysis, and wrote the manuscript. BG and AV validated, helped conceptualize, reviewed an early version of the manuscript, and provided resources. We are grateful to the University of Vienna for funding this fundamental science work[, making this work open access, ] and also acknowledge the Austrian Scientific Cluster for the allocation of computational resources.

The authors declare there are no conflicts of interest for this manuscript.

AI tools were used for correcting typos in the initial draft and to aid in constructing and validating the open data repository.

# Bibliography

Bony, S., and Coauthors, 2015: Clouds, Circulation and Climate Sensitivity. *Nat. Geosci.*, **8**, 261–268, https://doi.org/10.1038/ngeo2398.

Bourgeois, Q., A. M. L. Ekman, M. R. Igel, and R. Krejci, 2016: Ubiquity and Impact of Thin Mid-Level Clouds in the Tropics. *Nat. Commun.*, **7**, 12432, https://doi.org/10.1038/ncomms12432.

Hahn, C. J., W. B. Rossow, and S. G. Warren, 2001: ISCCP Cloud Properties Associated with Standard Cloud Types Identified in Individual Surface Observations. *J. Clim.*, **14**, 11–28, https://doi.org/10.1175/1520-0442(2001)014<0011:ICPAWS>2.0.CO;2.

van Heerwaarden, C. C., B. J. H. van Stratum, T. Heus, J. A. Gibbs, E. Fedorovich, and J. P. Mellado, 2017: MicroHH 1.0: A Computational Fluid Dynamics Code for Direct Numerical Simulation and Large-Eddy

Simulation of Atmospheric Boundary Layer Flows. *Geosci. Model Dev.*, **10**, 3145–3165, https://doi.org/10.5194/gmd-10-3145-2017.

Hogan, R. J., and A. Bozzo, 2018: A Flexible and Efficient Radiation Scheme for the ECMWF Model. *J. Adv. Model. Earth Syst.*, **10**, 1990–2008, https://doi.org/10.1029/2018MS001364.

Johnson, R. H., T. M. Rickenbach, S. A. Rutledge, P. E. Ciesielski, and W. H. Schubert, 1999: Trimodal Characteristics of Tropical Convection. *J. Clim.*, **12**, 2397–2418, https://doi.org/10.1175/1520-0442(1999)012<2397:TCOTC>2.0.CO;2.

Larson, V. E., A. J. Smith, M. J. Falk, K. E. Kotenberg, and J.-C. Golaz, 2006: What Determines Altocumulus Dissipation Time?. *J. Geophys. Res. Atmos.*, **111**, https://doi.org/10.1029/2005JD007002.

Li, J., J. Huang, K. Stamnes, T. Wang, Q. Lv, and H. Jin, 2015: A Global Survey of Cloud Overlap Based on CALIPSO and CloudSat Measurements. *Atmos. Chem. Phys.*, **15**, 519–536, https://doi.org/10.5194/acp-15-519-2015.

Liu, L., Y. Huang, and J. R. Gyakum, 2025: Clouds Reduce Downwelling Longwave Radiation over Land in a Warming Climate. *Nature*, **637**, 868–874, https://doi.org/10.1038/s41586-024-08323-x.

Luo, H., J. Quaas, and Y. Han, 2023: Examining Cloud Vertical Structure and Radiative Effects from Satellite Retrievals and Evaluation of CMIP6 Scenarios. *Atmos. Chem. Phys.*, **23**, 8169–8186, https://doi.org/10.5194/acp-23-8169-2023.

Lutsko, N. J., C. J. Wall, and B. Gasparini, 2026: An Analytical Model of the Lifecycle of Tropical Anvil Cloud Radiative Effects. *J. Geophys. Res. Atmos.*, **131**, e2025JD045070, https://doi.org/10.1029/2025JD045070.

McGraw, Z., B. Gasparini, W. Mol, and A. Voigt, 2026: Reconciling the Lack of a Robust Anvil Cloud Amount Response to Warming, https://doi.org/10.48550/arXiv.2609.00429.

Pan, Z., D. Rosenfeld, L. Zang, J. Yin, and F. Mao, 2025: Large Warming of Tropical Convective Anvils Masked by Their Underlying Clouds. *Remote Sens. Environ.*, **317**, 114524, https://doi.org/10.1016/j.rse.2024.114524.

Riihimaki, L. D., S. A. McFarlane, and J. M. Comstock, 2012: Climatology and Formation of Tropical Midlevel Clouds at the Darwin ARM Site. *J. Clim.*, **25**, 6835–6850, https://doi.org/10.1175/JCLI-D-11-00599.1.

Sassen, K., and Z. Wang, 2008: Classifying Clouds around the Globe with the CloudSat Radar: 1-Year of Results. *Geophys. Res. Lett.*, **35**, https://doi.org/10.1029/2007GL032591.

Seeley, J. T., N. Jeevanjee, W. Langhans, and D. M. Romps, 2019: Formation of Tropical Anvil Clouds by Slow Evaporation. *Geophys. Res. Lett.*, **46**, 492–501, https://doi.org/10.1029/2018GL080747.

Sokol, A. B., and D. L. Hartmann, 2022: Congestus Mode Invigoration by Convective Aggregation in Simulations of Radiative-Convective Equilibrium. *J. Adv. Model. Earth Syst.*, **14**, e2022MS003045, https://doi.org/10.1029/2022MS003045.

Spaulding-Astudillo, F. E., and J. L. Mitchell, 2025: Clear-Sky Convergence, Water Vapor Spectroscopy, and the Origin of Tropical Congestus Clouds. *AGU Adv.*, **6**, e2024AV001300, https://doi.org/10.1029/2024AV001300.

Stevens, B., and Coauthors, 2026: ORCESTRA: Organized Convection and EarthCARE Studies Over the Tropical Atlantic. *Tellus*, **79**, https://doi.org/10.16993/tellus.4123.

Stubenrauch, C. J., and Coauthors, 2013: Assessment of Global Cloud Datasets from Satellites: Project and Database Initiated by the GEWEX Radiation Panel. *Bull. Am. Meteorol. Soc.*, **94**, 1031–1049, https://doi.org/10.1175/BAMS-D-12-00117.1.

Sullivan, S., and B. Gasparini, 2024: Clouds and Radiatively Induced Circulations. *Clouds and Their Climatic Impact*.

Tanelli, S., S. L. Durden, E. Im, K. S. Pak, D. G. Reinke, P. Partain, J. M. Haynes, and R. T. Marchand, 2008: CloudSat's Cloud Profiling Radar After Two Years in Orbit: Performance, Calibration, and Processing. *IEEE Trans. Geosci. Remote Sens.*, **46**, 3560–3573, https://doi.org/10.1109/TGRS.2008.2002030.

Winker, D. M., and Coauthors, 2010: The CALIPSO Mission: A Global 3D View of Aerosols and Clouds. *Bull. Am. Meteorol. Soc.*, **91**, 1211–1230, https://doi.org/10.1175/2010BAMS3009.1.

Zhang, D., Z. Wang, and D. Liu, 2010: A Global View of Midlevel Liquid-Layer Topped Stratiform Cloud Distribution and Phase Partition from CALIPSO and CloudSat Measurements. *J. Geophys. Res. Atmos.*, **115**, https://doi.org/10.1029/2009JD012143.

# Supporting information

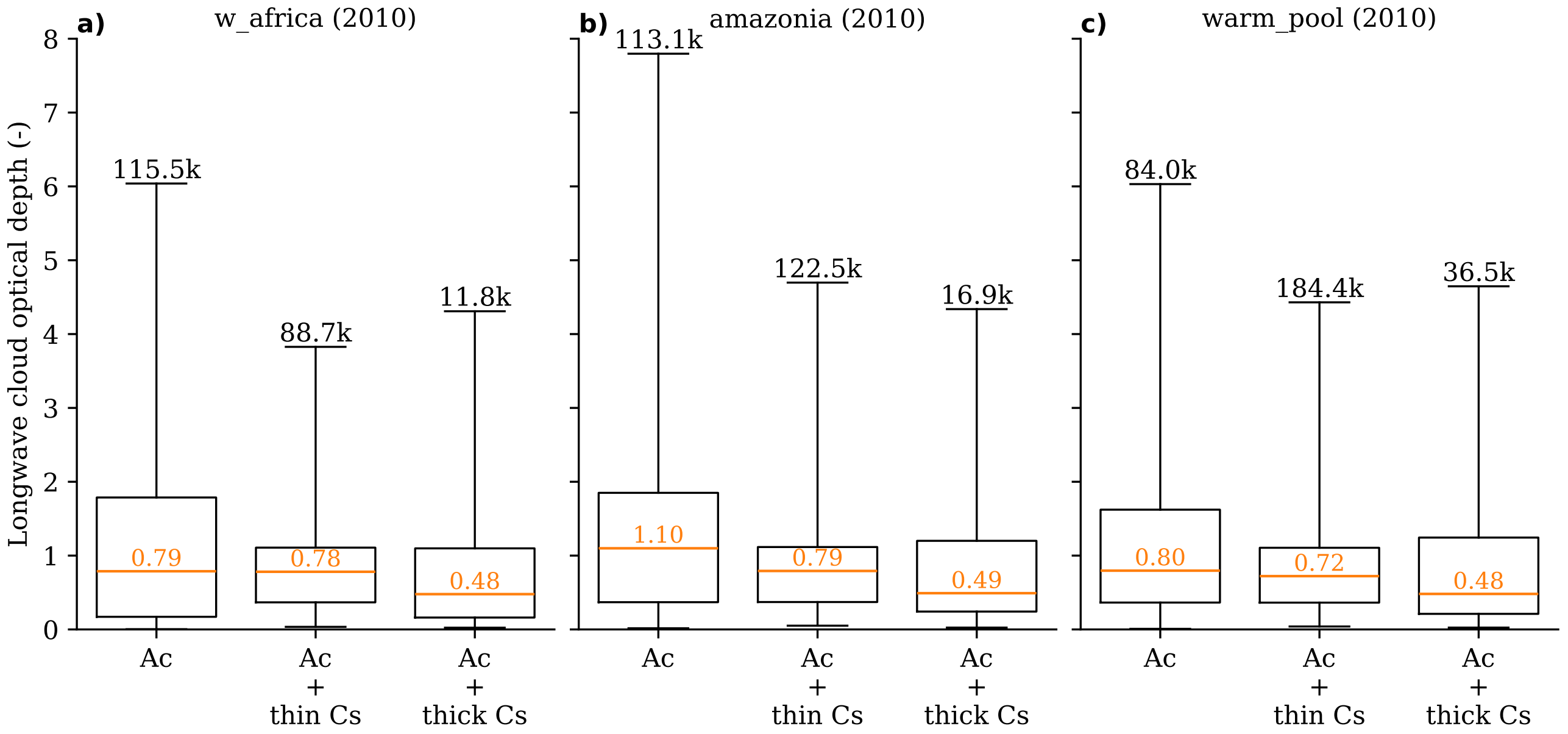

Figure S1: Cloud optical depth (broadband longwave) box plots derived from the CloudSat-CALIPSO data. Data is filtered for all Ac, Ac with thin Cs ($\tau < 0.5$), or Ac with thick Cs ($\tau \geq 0.5$). Ac here has a vertical depth $< 1000$ m. The box plots span the 5 to 95th percentile range, the boxes themselves are the 25 to 75th percentiles. The numbers are absolute counts of occurrences. Orange lines and text labels show the medians.

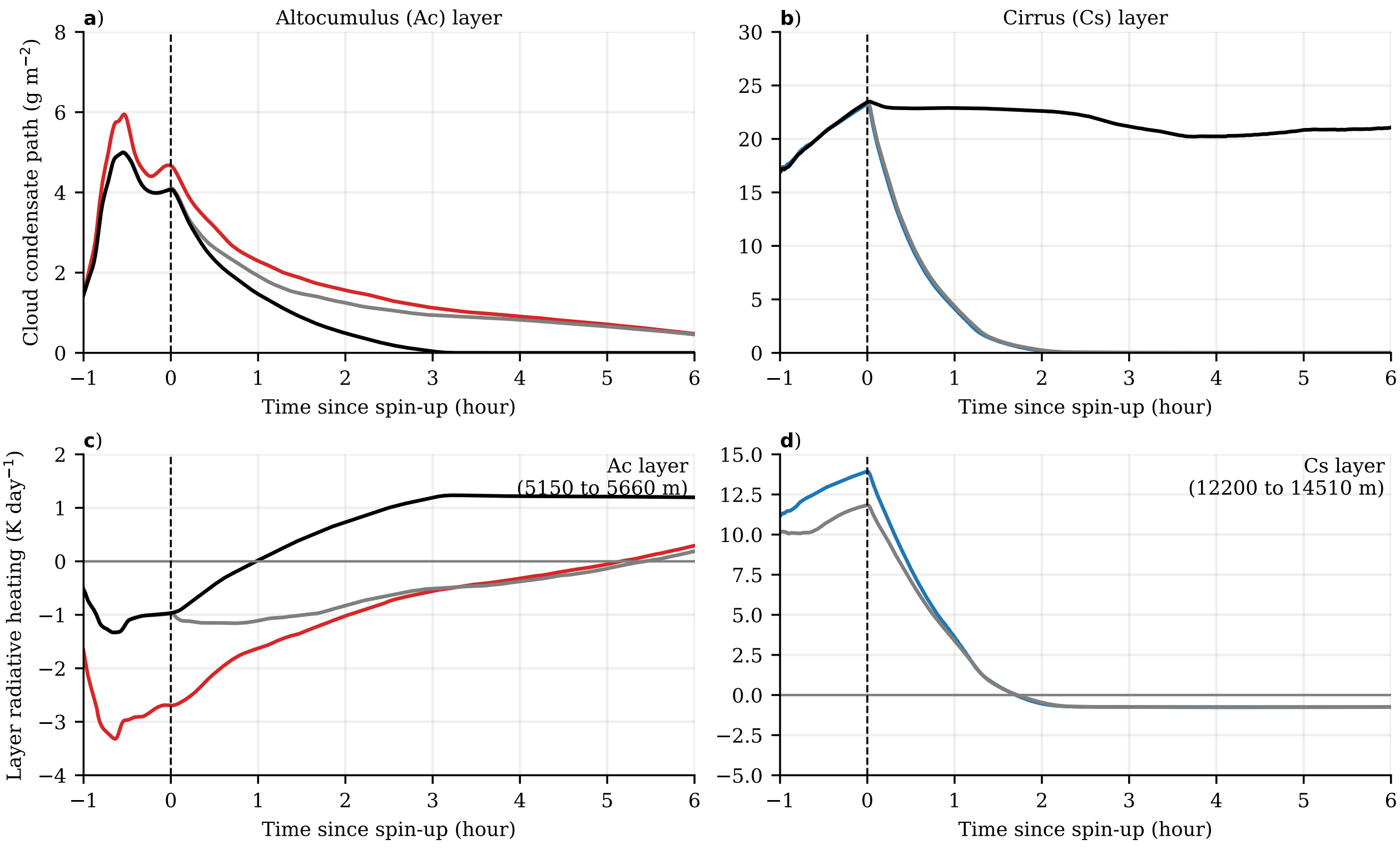

Figure S2: Same as Figure 3, but for the experiment with added daytime solar radiation.

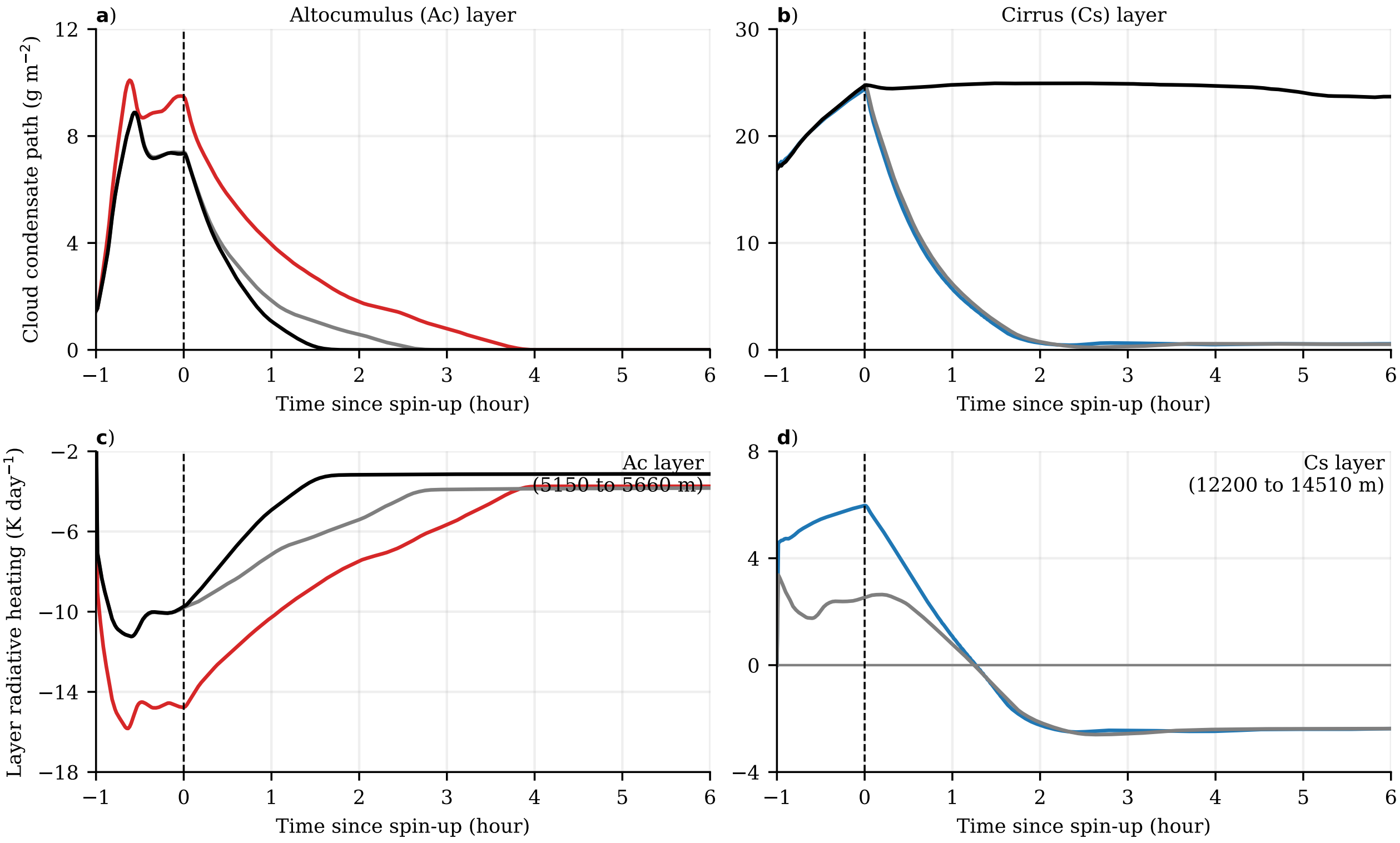

Figure S3: Same as Figure 3, but for the experiment with added subsidence.